\documentclass{iucr}
   \papertype{CP} 
    \journalcode{J}

\usepackage{xcolor}
\usepackage{amsmath}
\usepackage{graphicx}
\usepackage[utf8]{inputenc}

\begin{document}

\hyphenation{MATLAB}

%-------------------------------------------------------------------------
% The header of the paper
%-------------------------------------------------------------------------

\title{\textsc{GenL}: An extensible fitting program for Laue oscillations and whole pattern fitting}

\author{Anna L.}{Ravensburg}

\author{Johan}{Bylin}
\author{Vassilios}{Kapaklis}
\author{Gunnar K.}{P\'alsson}
%\aufn{These authors have contributed equally to the project.}

\aff{Department of Physics and Astronomy, Uppsala University, Box 516, SE-75120 Uppsala \country{Sweden}}

\maketitle

\begin{synopsis}
Supply a synopsis of the paper for inclusion in the Table of Contents.
\end{synopsis}

\begin{abstract}
\textsc{GenL} is a flexible program that can be used to simulate and/or fit X-ray reflectivity and X-ray diffraction data from epitaxial thin films exhibiting, for example, Laue oscillations. It utilizes a differential evolution within a genetic algorithm for fitting data and uses a modular approach based on either the kinematic theory of diffraction or the dynamic theory. Effects of polarization, absorption, the Lorentz factor, as well as instrumental resolution and lattice vibrations are taken into account. Useful parameters that can be extracted after fitting include atomic interplanar spacings, number of coherently scattering atomic planes, strain profiles along the film thickness, and crystal roughness. The program has been developed in \textsc{MATLAB} and employs a graphical user interface. The deployment strategy is twofold, whereby the software can either be obtained in source code form and executed within the \textsc{MATLAB} environment, or as a pre-compiled binary for those who prefer not to run it within \textsc{MATLAB}. Finally, \textsc{GenL} can be easily extended to simulate multilayered film systems, superlattices, and films with atomic steps. The program is released under the GNU General Public License.
\end{abstract}

%-------------------------------------------------------------------------
% The main body of the paper
%-------------------------------------------------------------------------

%************Introduction************
\section{Introduction and motivation}\vspace{0.5mm}

High crystal quality epitaxial thin films are intriguing model systems. Low defect density and minor crystal misorientation allow for a direct and easy comparison to theoretical predictions \cite{Neugebauer2013}. To experimentally verify the crystal quality of epitaxial thin films, the presence of Laue oscillations in X-ray diffraction (XRD) patterns is regarded as a key indication of high crystallinity \cite{Miller2022}. Laue oscillations are only observable if the coherence length of the epitaxial layer is on the order of the film thickness, and the latter is smaller than the longitudinal coherence length of the X-ray photons. The occurrence of Laue oscillations is often used to qualitatively characterize samples, associating their presence with high crystallinity and low defect density  \cite{Krauss2008, Shu2020, Song2020, Kamigaki1990, DeTheresa2007}, thickness uniformity \cite{Naito2001, Xu2017, Hu2013}, smooth and flat interface layering \cite{Abe2002, Xu2017, Miyadera2021, Radu2006}, and film homogeneity \cite{Forst2015}. Nevertheless, a quantitative analysis of the Laue oscillations, modeling their angular spacing, intensity decay, and asymmetry around their respective Bragg peak can yield additional structural insights \cite{Fullerton1992, Stierle1993, Komar2017, Miller2022}.

For instance, the characteristic period of the Laue oscillations is related to the number of coherently scattering layers, and hence, to the coherent thickness of the layer. However, the thickness determined via the separation of Laue oscillations can deviate from the thickness determined by the spacing of the Kiessig fringes found in reflectivity measurements \cite{Kiessig1931}. A possible origin of this may be the presence of excess material at an interface \cite{Miller2022} with a similar electron density as the film but with a different crystal structure. This region adds to the total film thickness determined via the spacing of the Kiessig fringes but it would not affect the thickness determined via the Laue oscillations as the mismatch in crystal structure terminates the crystal coherence length of the film. The intensity decay of the Laue oscillations can be understood in terms of the roughness of the layer, i.e., the distribution of out-of-plane thickness that contributes to coherent scattering. Moreover, an asymmetry of the Laue oscillation intensity on either side of the Bragg peak is frequently observed \cite{Miller2022} and often associated with out-of-plane strain \cite{Vartanyants2000, Robinson2001, Kastle2002, Komar2017}, or other factors~\cite{Miller2022}. Numerous powerful open-source program codes exist for simulating and fitting of diffraction patterns including Laue oscillations, e.g., \textsc{SUPREX} \cite{Fullerton1992}, \textsc{GenX} \cite{Bjorck2007, Glavic_Bjorck_2022}, \textsc{CADEM} \cite{Komar2017}, or \textsc{InteractiveXRDFit} \cite{Lichtensteiger2018} alongside with commercially available software, e.g., \textsc{BedeREFS} \cite{Wormington1999} or \textsc{Leptos} \cite{Ulyanenkov2004}. However, the focus of most of them lies mainly on the fitting of superlattice peaks and not on the detailed analysis of the Laue oscillations and their shape.
The open-source software \textsc{SUPREX} is designed for superconductor superlattices and written in Fortran and Turbo Pascal \cite{Fullerton1992} with a non-trivial extension to other materials \cite{Komar2017}.
\textsc{GenX} is designed to fit X-ray reflectivity patterns \cite{Bjorck2007}, being primarily sensitive to electron density profiles along the z-axis. It is possible to adjust it in order to fit high-angle scattering data. Though, reconfiguring the program for these settings is not effortless and detailed analysis of the shape of the oscillations is difficult to implement.
\textsc{CADEM} on the other hand, which is written in \textsc{Matlab}, can be used to simulate scattering patterns without a fitting option and without the inclusion of crystal roughness.
Finally, \textsc{InteractiveXRDFit}, a \textsc{Matlab} program designed for oxide thin films and heterostructures, utilizes an intermediate approach between the two limits of the kinematic and dynamic X-ray diffraction theories to calculate, but not fit, the diffracted intensity \cite{Lichtensteiger2018}.
Besides these, Monte Carlo simulations can be used to reproduce experimental diffraction patterns, however, their application is hindered by their high degree of complexity for modeling realistic sample structures \cite{Komar2017}.

Hence, we identified the need for \textsc{GenL}, an open source and easy to adapt Laue oscillation fitting program, which is able to not only simulate but also fit experimental diffraction patterns of epitaxial thin films.
In \textsc{GenL}, a representative diffraction pattern, which includes all the aforementioned parameters influencing the shape of the measured oscillations, is simulated and fitted to experimental data.

The associated parameters are tuned via a differential evolution algorithm \cite{Storn1997} until a minimum difference of the figure of merit between the data and simulation has been obtained. 

In this work, we first present the theoretical background for the calculations of the diffraction patterns used in \textsc{GenL}.
We then illustrate how the program can be used to fit experimental data and to analyze the sample layering and crystal structure, with the option to include roughness as well as strain profiles and defects.
Finally, we show a few examples of the versatile applications of \textsc{GenL}.

%************Theoretical background************
\section{Theoretical background}\label{sec:theory}\vspace{0.5mm}

X-ray diffraction can be used to determine the separation between crystallographic planes $d_{hkl}$ via Bragg's law \cite{Bragg1913} :
% ---------------
\begin{equation}
\label{equ:bragg}
2d_{hkl}\sin(\theta_{hkl}) = \lambda,
\end{equation}
% ---------------
where $\lambda$ is the X-ray wavelength and $\theta _{hkl}$ is the angle at which constructive interference occurs between sets of planes associated with the Miller indices $hkl$.
Equivalently, the conditions for Bragg reflections are fulfilled when the photon's momentum transfer wave vector $\vec{Q}$ coincides with the reciprocal crystal lattice point represented by the vector $\vec{G}_{hkl}$.
In a symmetric $\theta$--$2\theta$ scan, the goniometer is driven so that the detector moves through $2\theta$ while the incident angle remains $\theta = (2\theta)/2$. Under this condition, the momentum-transfer vector varies only in the direction perpendicular to the lattice planes being probed, assuming proper alignment and correction for miscut and other systematic deviations.
Its modulus can then be written as:
% ---------------
\begin{equation}
\label{equ:q}
Q = \frac{4\pi\sin(\theta)}{\lambda}.
\end{equation}

\subsection{Kinematic approach}
Assuming the reciprocal lattice vector is oriented parallel to $\vec{Q}$, and referring to this direction as $z$ (with $x$ being perpendicular to $z$ and in the scattering plane), the intensity $I(Q)$ diffracted from the crystal can be written in the kinematic approximation as:
% ---------------
\begin{equation}
\label{equ:structureFactor}
I(Q) = \Big|\sum_{n=1}^{N_\text{L}} f_n(Q)e^{iQz_n}e^{-B_n \left(\frac{Q}{4\pi}\right)}\Big|^2,
\end{equation}
% ---------------
where $N_L$ corresponds to the total number of coherently scattering lattice planes within the coherence volume of the X-rays, $z_n$ is the position of the $n^{\rm th}$ atomic plane in the thin film stack (taking $z_1=0$ in the center of the first layer above the substrate interface), and $e^{-B_n\left(\frac{Q}{4\pi}\right)^2}$ is the Debye-Waller factor of layer $n$. This expression is generalized in the code to an arbitrary stack of layers, including a substrate, with spacing $d_i$ between layers, where $i$ is an interface index.

The element specific Debye-Waller pre-factor $B$ includes the contribution of the time-averaged displacement of atoms due to lattice vibrations.
The dispersion corrected and $Q$-dependent form factor $f(Q)$ is given by \cite{Wilson1993, Henke1993}:
% ---------------
\begin{equation}
\label{equ:formFactor}
f(Q) = f_\text{0}(Q) + \Delta f' + i\Delta f'',
\end{equation}
% ---------------
along with the anomalous $\Delta f'$ and absorption $\Delta f''$ corrections of Brennan and Cowan~\cite{Brennan1992}, whereas $f_0(Q)$ is commonly expressed in its parameterized form \cite{Waasmaier1995}:
% ---------------
\begin{equation}
\label{equ:nthformFactor}
f_\text{0}(Q) = \sum_{i=1}^{5} a_\text{i}e^{-b_\text{i}\left(\frac{Q}{4\pi}\right)^2} + c,
\end{equation}
% ---------------
where $a$, $b$, $c$, are the form factor parametrization coefficients of a particular atomic species.
For an alloy, $f(Q)$ is a mole fraction weighted sum of the form factors of the elements in the alloy (with the possibility of having different form factors for different oxidation states).

Several other factors need to be considered \cite{Warren1990,Fullerton1992} if the diffracted intensity is to be compared to the experimentally observed one.
The first factor $P(\theta)$ includes the inherent polarization dependence of the scattering of an X-ray with an electron, as well as the change in polarization when passing through a monochromator.
The second factor $A(\theta)$ accounts for the absorption of X-rays along the angle dependent path lengths of the incident and diffracted rays.
The third factor is the Lorentz factor $L(\theta)$ which takes into account deviations from a perfect instrument and the mosaic nature of the crystal.
The experimentally observed intensity is then given by:
% ---------------
\begin{equation}
\label{equ:intensity}
I_{\rm{obs}}(Q) = I(Q) \cdot P(\theta) \cdot A(\theta) \cdot L(\theta).
\end{equation}
% --------------
Further details on origin and derivation of these factors can be found elsewhere \cite{Buerger1945, Warren1990, Fewster1996, Birkholz2005}, but for the convenience of the reader, we give a brief overview of their importance here. 

The polarization factor $P(\theta)$ depends on whether a monochromator is used and whether it is located on the incidence side or on the detector side.
On the incidence side it takes the form \cite{Azaroff1955}:
% ---------------
\begin{subequations}
\label{equ:polarized_i}
\begin{align}
P(\theta) = \frac{1 + \cos^2(2\theta) \cos^2(2\theta_{\rm{M}})}{1 + \cos^2(2\theta_{\rm{M}})}\\
\intertext{while on the detector side \cite{Yao1982}:}
\label{equ:polarized_d}
P(\theta) = \frac{1 + \cos^2(2\theta) \cos^2(2\theta_{\rm{M}})}{2},
\end{align}
\end{subequations}
% --------------
where $\theta_{\rm{M}}$ is the diffraction angle of the monochromator.
If no monochromators are being used, then $\theta_{\rm{M}}$ is equal to zero.

The absorption factor $A(\theta)$ corrects for the angle-dependent attenuation of X-rays in the sample.
For a single layer measured using a $\theta/2\theta$ geometry, the absorption factor, corresponding to the effective combined probe and escape depth of the sample, is equal to:
% ---------------
\begin{equation}
\label{equ:absorption}
A(\theta) = \frac{1}{2\mu}\Big(1 - e^{-\frac{2\mu t}{\sin \theta}}\Big),
\end{equation}
% --------------
where $t$ is the thickness of the film.
We assume all extinction effects are negligible.
The linear attenuation coefficient $\mu$ can be computed as the product of the atomic density $\rho$ (number of formula units per volume) and the photoabsorption cross-section of the sample, which is included in the form factor correction $\Delta f''$.
The linear attenuation coefficient is then written as:
% ---------------
\begin{equation}
\label{equ:mu}
%\mu = 2r_e\rho \lambda \, {\rm Im}(f),
\mu = 2r_e\rho \lambda \sum\limits_{i=1}^m c_i \Delta f''_i,
\end{equation}
% --------------
where $r_e$ is the electron radius, $m$ is the number of elements in the alloy, and $c_i$ is the mole fraction of element $i$.

Finally, as an option in the program, the Lorentz factor $L(\theta)$ captures the angular dependent changes of the intensity, due to the combination of monochromator imperfections, beam divergence, finite detection resolution, and mosaicity in the sample and depends on the measurement geometry.
Further details can be found elsewhere \cite{Warren1990}.
The combined change in intensity can be corrected via:
% ---------------
\begin{equation}
\label{equ:lorentz}
L(\theta) = \frac{1}{\sin(2\theta)}.
\end{equation}
% --------------
\section{Dynamical approach}
The dynamic version of the code is an exact calculation of the X-ray scattering using Hol\'y and Fewster’s formulation~\cite{Holy_Fewster}. The essential feature is a matrix propagation using Parratt’s algorithm of the (complex) electron density through the entire stack, including any substrate. In contrast to the algorithm in \textsc{GenX} where slab models are used for each physical layer, here every unit cell is further divided into a number of slabs (typically 100). The algorithm is therefore expected to be much slower than a typical reflectivity calculation, but through a number of optimization steps the run time has been cut by several orders of magnitude compared to a naive implementation, thereby rendering these calculations practical. However, the kinematic approach is still faster and is recommended for initial fits to the data. Furthermore, matrix propagation diverges when large number of slabs are used, but is avoided here by a normalization procedure. Our implementation differs from Hol\'y and Fewster’s in that in their work the electron density is calculated in reciprocal space, whereas we calculate the density in real space. This is to allow for the possibility of strain and roughness, which are easier to conceptualize in real space. 

The electron density is calculated~\cite{Holy_Fewster} if the type of atoms and their coordinates are known, by overlaying the inverse Fourier transforms of the atomic form factors (including dispersion corrections, absorption and Debye-Waller factors) and dividing by the in-plane area of the unit cell. The average electron density is calculated on a grid of points specified by the user, as well as the the limits of integration of the form factors. Strain is implemented by moving the coordinates prior to the transform. At present, a linear strain profile is included from one or both sides of each layer. Crystal roughness is implemented by a Gaussian weighting procedure of calculations of slabs with a different number of unit cells (truncated at three standard deviations). Monolayers can be implemented as separate unit cells by the user. Finally, a layer of vacuum (or air) is added to both sides of the stack. Once the electron density per unit length has been obtained, the diffraction pattern is calculated by matrix propagation, taking the polarization of the incoming X-ray into account and squaring the reflectivity coefficient at the topmost layer. Thus, reflectivity, absorption, and transmission can be calculated to any precision from an arbitrary stack of layers including the effects of strain and roughness. This includes effects of refraction, Darwin broadening, and total reflection; all effects that are not present in the kinematic approach. We note that the density profile generated by \textsc{GenL} cannot be fed into \textsc{GenX} to calculate the diffraction pattern, owing to the divergences and prohibitively slow calculation. To simulate unpolarized light we perform both the dynamical calculation for s and p polarization and combine them. For the case where the monochromator is located on the incidence side, we obtain:
\begin{equation}
I_{\rm{obs}}^{dyn}(Q) = \frac{I_{\rm{s}}^{dyn}(Q) + \cos(2\theta_{\rm{M}})^2I_{\rm{p}}^{dyn}(Q)}{1 + \cos(2\theta_{\rm{M}})^2}
\end{equation} 
and a similar expression for when the monochromator is on the detector side. We have omitted the Laue factor for the dynamical case because it diverges at low angles.

\begin{figure}
\centering
\includegraphics[width=0.95\linewidth]{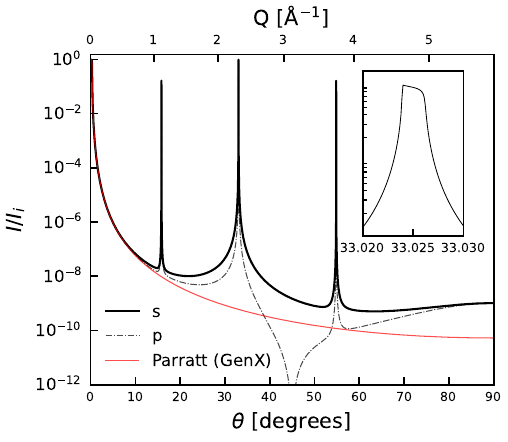}
\caption{Simulated X-ray diffraction pattern of a GaAs crystal using the dynamic formalism, assuming Cu K$\alpha_1$ radiation. Curves are shown for both the s and p polarizations, along with a comparison to the output of \textsc{GenX} for the same structure. The inset depicts a detailed view of the strongest diffraction peak for s polarization and the characteristic Darwin shape.} 
\label{fig:GaAs}\hfill
\end{figure}

\begin{figure}
\centering
\includegraphics[width=0.95\linewidth]{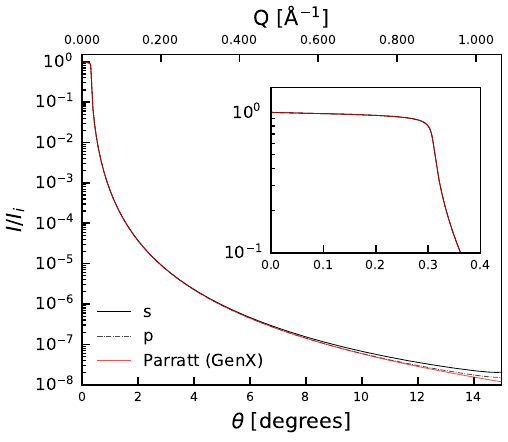}
\caption{Comparison of simulated reflectivity curves for a GaAs substrate, assuming Cu K$\alpha_1$ radiation. The curves have been calculated using the dynamical approach and are compared with the Parratt formalism as implemented in \textsc{GenX}. The inset provides a detailed comparison for angles below the total reflection.
} 
\label{fig:GaAs_XRR}\hfill
\end{figure}

Fig. \ref{fig:GaAs} depicts the diffraction pattern using the dynamical approach for a bulk single crystal of GaAs with vacuum on either side. The reflectivity region exactly corresponds to the one calculated from \textsc{GenX} as shown in Fig. \ref{fig:GaAs_XRR} and the diffraction pattern exhibits Darwin widths as calculated by Hol\'y and Fewster (inset of Fig. \ref{fig:GaAs}). Both s and p polarizations are shown in the figure, in excellent agreement with the calculations in [Ref.~\cite{Holy_Fewster}]. Note that the dynamical results deviate from the results from \textsc{GenX} at larger scattering angles, highlighting the potential importance of including diffraction effects when fitting reflectivity.

% ---------------
% ---------------

Moreover, electron density obtained from density functional theory or other codes can be used as direct input into the code and the exact diffraction pattern can be calculated, which can be useful when studying subtle changes to the electron density associated with for example Jahn-Teller distortions or crystal field splitting, and magnetic effects, which are not captured by the use of tabulated form factors.

% ---------------
% ---------------

%************GenL************
\section{\textsc{GenL}}\vspace{0.5mm}

In \textsc{GenL}, a diffraction pattern is calculated based on the stated equations.
However, to model X-ray scattering intensity measured in a non-ideal diffractometer setup of a non-perfect single crystal, additional intensity contributions are included in the calculation.
Instrumental peak broadening is accounted for by a Gaussian convolution.
Moreover, the background intensity $I_{\rm bkg}(Q)$ originating from thermal diffuse scattering, fluorescence radiation, Compton scattering, air scattering, the detector dark current, and coherent diffuse scattering from defects needs to be included.
In the close vicinity of a peak, assuming a linear dependence often allows for a reliable modeling of $I_{\rm bkg}(Q)$ \cite{Birkholz2005}:
% ---------------
\begin{subequations}
\label{equ:background_l}
\begin{align}
I_{\rm bkg}(Q) = a Q + b.\\
\intertext{However, for certain samples, which, e.g., include amorphous layers (for example protective cap layers), a more complex polynomial modeling might be needed, such that:}
\label{equ:background_p}
I_{\rm bkg}(Q) = c (a + Q)^2 + b,
\end{align}
\end{subequations}
% --------------
where $a$, $b$, and $c$ are constants.
Both options are available to choose from in \textsc{GenL}.
Furthermore, several choices are available for inclusion of the substrate in the intensity calculation in \textsc{GenL} for the kinematic case,  $I_{\rm sub}(\theta)$.
The first choice is to model the substrate Bragg peak with a Lorentzian function following:
% ---------------
\begin{equation}
\label{equ:lorentzian}
I_{\rm sub}(\theta) = \frac{I_{\rm 0,sub}}{\pi} \left[\frac{w_{\rm sub}}{(2\theta-x_{\rm 0,sub})^2+w_{\rm sub}^2}\right],
\end{equation}
% --------------
where $I_{\rm 0,sub}(\theta)$ is the substrate peak intensity, $w_{\rm sub}$ is the half width at half maximum of the peak and $x_{\rm 0,sub}$ is the expected peak position in $2\theta$.
However, for the case that the Bragg peak positions from film and substrate lie close to each other, scattering from film and substrate will interfere, which can cause an asymmetry in the Laue oscillations as well as an intensity shift of the Bragg peak position \cite{Robinson1988}.
An accurate calculation of the scattering pattern hence requires to add the scattering amplitudes from film and substrate prior to squaring.
Using the \textsc{GenL} command line version the substrate can be included as any other layer, either in the kinematic or exact dynamic approach.
In addition, in \textsc{GenL}, out-of-plane strain in the thin film as well as layer roughness can be taken into account.
Strain in a single layer is modeled in the code by displacing the atoms in the calculated one-dimensional stack according to a certain strain profile, which can be either exponential or linear.
For the exponential case the displacement $\epsilon_\text{n}$ of the $n^{\mathrm{th}}$ atom from its position in an unstrained lattice \cite{Fullerton1992} is given by:
% ---------------
\begin{equation}
\label{equ:strain}
\epsilon_\text{n} = e^{-\alpha \cdot d_\text{0} \cdot n},
\end{equation}
% ---------------
where $d_\text{0}$ denotes the out-of-plane distance between the atomic planes for the unstrained case and $\alpha$ is a fitting parameter. For the linear strain profile, both the slope and the extent into the layer can be fitted.
In the program we consider strain originating from the substrate/film interface ($\alpha_1$) and strain originating from an additional film/capping interface ($\alpha_2$). The latter is not relevant for sole single layers without capping. The respective induced strain can either be tensile or compressive, resulting in the addition or subtraction of $\epsilon_\text{n}$ to/from the position in an unstrained lattice, respectively.
The layer roughness $\sigma$ is included in \textsc{GenL} employing:
% ---------------
\begin{equation}
\label{equ:rough}
I \propto \sum_{N \in A} \frac{1}{\sqrt{2\pi}\sigma} e^{-\frac{(N-N_L)^2}{2\sigma^2}},
\end{equation}
% ---------------
where the set $A$ are all integers in a $3\sigma$ interval around $N_L$, i.e., $[N_L-3\sigma,N_L+3\sigma]$ \cite{Fewster1996}.
In the command line version of \textsc{GenL}, any other model for strain and layer roughness can be manually included, e.g., strain due to charge density waves \cite{Singer2016} or roughness based on binomial fluctuations \cite{Miceli1992}.

For a single layer, using the kinematic formulation we have implemented a graphical user interface, whereby the theoretical X-ray scattering pattern calculated in \textsc{GenL} is fitted to the measured intensity using 13 parameters:
\begin{itemize}
    \item the out-of-plane interplanar spacing $d$
    \item the number of coherently scattering planes $N$
    \item resolution and scale factor for the Gaussian convolution
    \item parameters modeling the background intensity $a$, $b$, and $c$
    \item parameters modeling a substrate peak $I_{\rm 0,sub}$, $w_{\rm sub}$, $x_{\rm 0,sub}$
    \item the strain parameters $\alpha_1$ and $\alpha_2$
    \item the roughness parameter $\sigma$
\end{itemize}

The fitting in \textsc{GenL} utilizes differential evolution within a genetic algorithm \cite{Storn1997} like the one used in \textsc{GenX} \cite{Bjorck2007, Glavic_Bjorck_2022}.
The algorithm first generates parameter vectors. The population of parameter vectors (parent population) is then changed based on a certain process to create a new population with a better fit compared to the previous one until the stopping criterion -- in \textsc{GenL} the maximum number of iterations -- is fulfilled. 
Further details on the used algorithm as well as on the population generation can be found elsewhere \cite{Bjorck2007}.
The genetic algorithm can be tuned in the command line version of \textsc{GenL} in terms of the population size and crossover factor.

\textsc{GenL} is similar to \textsc{GenX} as it also offers a graphical user interface (GUI), both are open-source and can be customized individually for specific applications.
Furthermore, both have embedded libraries/interfaces with databases to fetch material and scattering parameters.
However, in \textsc{GenX} the layering in a thin film stack is simulated, whereas \textsc{GenL} simulates the atomic ordering.
It provides deeper insights into the atomic structures and crystal properties, which \textsc{GenX} fits cannot do.
In contrast to \textsc{GenX}, \textsc{GenL} is not applicable to all kinds of layers, e.g., amorphous layers. In cases where information about interatomic spacing and coordination is required, recently developed tools designed to determine pair distribution functions of thin films using laboratory-based X-ray sources can be utilized \cite{one_shot_PDFs}.

\begin{figure}
\centering
\includegraphics[width=0.95\linewidth]{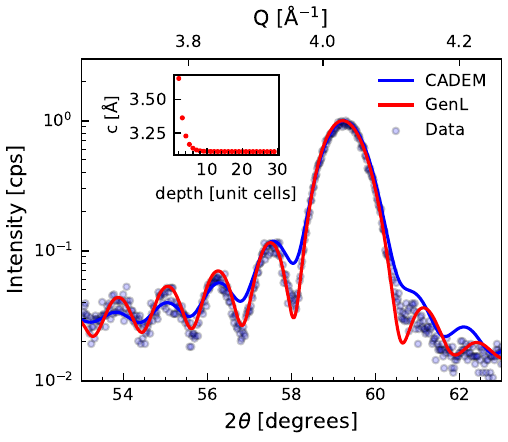}
\caption{X-ray diffraction pattern of a 105~{\AA} thick V layer on a MgO~($001$) substrate. The pattern was previously published in \cite{Komar2017} with a simulation of the diffraction pattern created with \textsc{CADEM}. This simulation is displayed in blue. A fit of the data with \textsc{GenL} is shown in red. The corresponding strain profile, i.e., the out-of-plane lattice parameter $c$ as a function of the number of unit cells counting from the substrate interface, is displayed as an inset.} 
\label{fig:CADEM}\hfill
\end{figure}

An example of a \textsc{GenL} fit of the scattering pattern of an epitaxial vanadium thin film is displayed in Fig.~\ref{fig:CADEM}.
Based on XRR measurements, the thickness of the V layer is 105~{\AA}.
The fit takes the tensile out-of-plane strain due to the negative lattice mismatch between the vanadium and the MgO~($001$) substrate into account.
The calculated, exponentially decaying strain profile is shown as an inset.
The \textsc{CADEM} simulation for the same data set published by Komar {\it et al.} \cite{Komar2017} is displayed as a blue line.
Also the \textsc{CADEM} simulation includes a strain profile \cite{Komar2017}, which is in qualitative agreement with the strain profile presented here.
The magnitude of the exponential decay in out-of-plane lattice parameter $c$ as a function of position is larger for the fit with \textsc{GenL}, in particular for depths in the sample corresponding to the first two unit cells.
The drastic change in $c$ is, hence, calculated to occur only close to the substrate/V interface and might include contributions from interface roughness or atomic steps, which may contribute to an asymmetric decay of the oscillations but are not considered here.
The over-estimation of the tetragonal unit cell distortion at the interface is therefore considered to be an artifact.
Numeric values of fitted strain profiles should be analyzed carefully and may be considered only a qualitative display if important contributions to the scattering intensity, especially at interfaces, have been neglected in the selected model for a certain sample.
The \textsc{CADEM} simulation corresponds to a coherent thickness of 84~{\AA}.
Based on the \textsc{GenL} fit, the coherent thickness of this V film is 86.3~{\AA}.
Komar {\it et al.} \cite{Komar2017} attribute this difference to the by XRR determined thickness to a partly oxidized surface with similar electron density but different crystal structure. 
We see that we can account for the asymmetry of the profile by including a strain profile combined with crystal roughness.

% ---------------
% ---------------

\subsection{Technical details}

\textsc{GenL} has been developed in the high-level programming language \textsc{MATLAB} (The MathWorks Inc., Natick, MA, USA).
\textsc{MATLAB} is available for numerous operating systems, namely Windows, macOS, and Linux/Unix. 
The \textsc{GenL} program can either run from the command line after adjusting the input file in a text editor, or the \textsc{GenL} GUI can be used. The GUI currently supports fitting a single layer using the kinematic approximation whereas the command line version supports any structure and either the kinematic or the dynamic approach.
Since the code is rather compact, an adaptation to other programming languages such as \textsc{Python}, \textsc{C}, \textsc{C++}, or \textsc{Java} is straightforward.
However, any alteration of the GUI requires a deeper understanding of the \textsc{MATLAB} programming language.
The command line version of the program was tested on the following \textsc{MATLAB} versions: R2019b, R2020b, R2022b, R2023a, and R2023b.
No additional toolbox installations are required.
The \textsc{GenL} GUI runs correctly in \textsc{MATLAB} versions published in 2022 or later.
In older versions, the graphical display in the app might be distorted.
Possible distortions of the graphical display in all versions can be avoided if the program is running on one screen only. 

% ---------------
\begin{figure}
  \centering
  \includegraphics[width=0.95\linewidth]{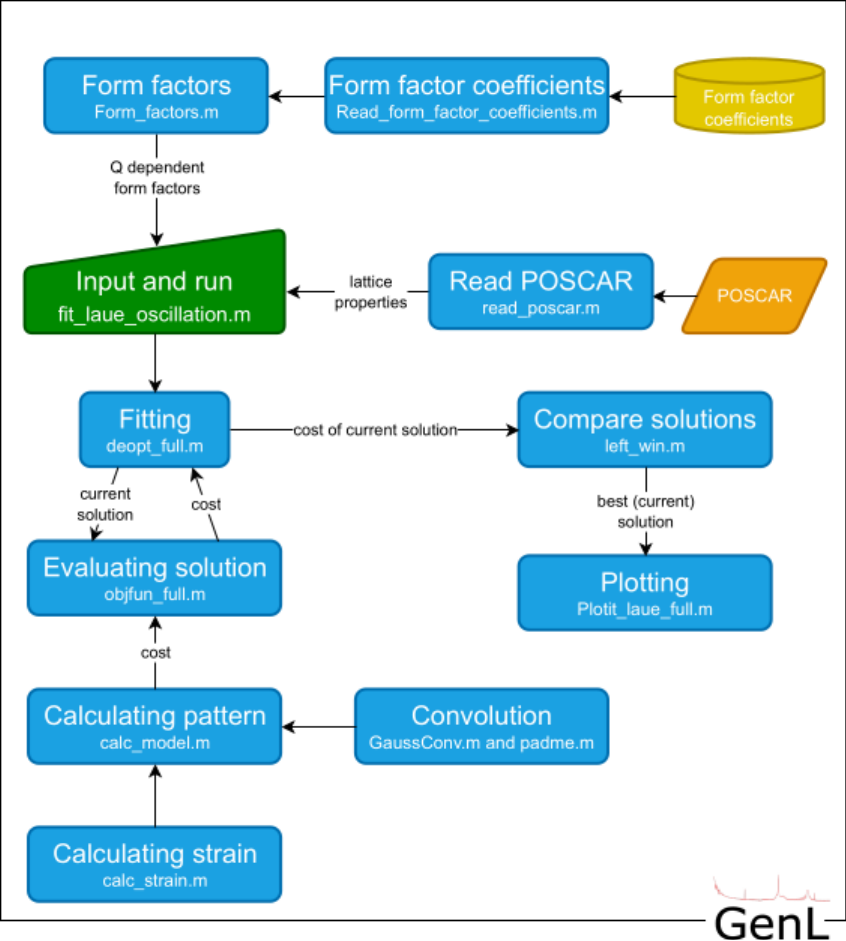}
  \caption{Flow chart of the different modules of the program. Corresponding functions/folders in \textsc{GenL} are displayed in the bottom of the respective boxes.}
  \label{fig:GenL}\hfill
\end{figure}
% ---------------

\subsection{Using \textsc{GenL}}

For running \textsc{GenL} from the command line, the user is required to edit the input file "{\it fit\textunderscore laue\textunderscore oscillations.m}".
This input file is displayed as a green trapeze shaped box in the flow chart in Fig.~\ref{fig:GenL}.
First, file and path of the measurement as well as important measurement parameters need to be specified.
Second, material specific parameters (atomic number $Z$, density, chemical composition, the Debye-Waller pre-factor), sample parameters (layer thickness, expected Bragg angle/interplanar spacing), and instrument parameters (wavelength, setup, measurement time, amplitude, resolution, and background level) need to be added.
For a limited number of elements, the element-specific Debye-Waller pre-factor can be extracted from a database \cite{Sears1991} as well, or be put in manually.
Furthermore, three features can be included in the fitting process of the program: strain, roughness, and a potentially present peak originating from a single-crystalline substrate can be accounted for by adding a Lorentzian intensity contribution at the respective Bragg angle.
If any of these features are included, their respective parameters need to be specified in the code as well.
Finally, for all upper and lower bounds for the fitting parameters are constructed and can be adjusted as necessary.
Running the input file will start the fitting process, which is further described in the next section.

\subsection{Program organisation}

A flow chart of all \textsc{GenL} modules including their function names is displayed in Fig.~\ref{fig:GenL}.
When \textsc{GenL} is executed, form factors are constructed following equations \ref{equ:formFactor} and \ref{equ:nthformFactor}. 
In the input file a fitting function is called with start values as well as upper and lower boundaries for all fitting parameters.
The fitting function creates a current solution for all parameters, based on which a scattering pattern is calculated according to the equations in section \ref{sec:theory} in the evaluation function.
The evaluation function compares the theoretical intensity profile and the data input and calculates the goodness of the fit for the current solution using a logarithmic figure of merit.
It is straightforward to add other figures of merit to the code.
In the fitting function the figures of merit of solutions are compared.
The fitting function and the function that compares different solutions were adapted from the work of Storn {\it et al.} \cite{Storn1997}.
If the current solution is an improvement compared to previous solutions, the fit is plotted and the corresponding parameters are printed out in the command window.
A maximum number of iterations is specified by the user in the input file.

\subsection{Graphical user interface}

\begin{figure}
\centering
\includegraphics[width=0.95\linewidth]{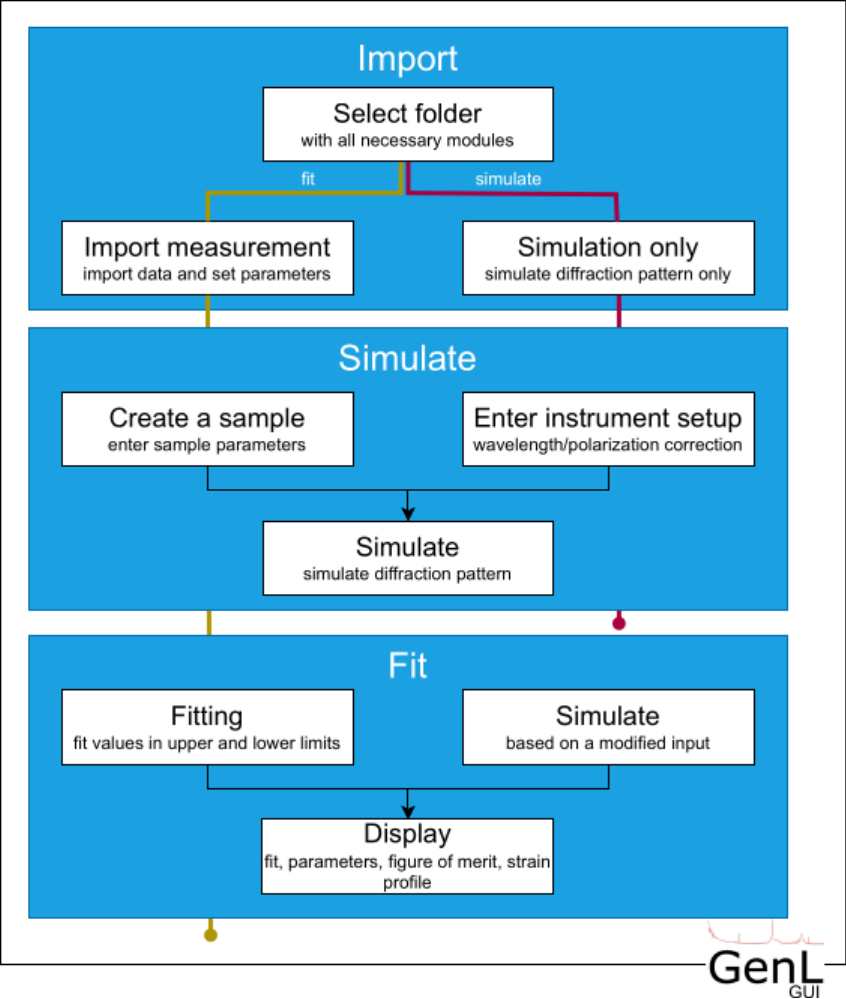}
\caption{Flow chart of the \textsc{GenL} GUI. Blue boxes correspond to different tabs, white boxes represent processes in the tabs. There are two possible paths using the program: either first uploading data, simulating a diffraction pattern, and fitting the data (yellow path), or only simulating a diffraction pattern based on the choice of input parameters (pink path).} 
\label{fig:GenL_GUI}\hfill
\end{figure}

\textsc{GenL} can also be used via the \textsc{GenL} GUI.
The program can be executed within the \textsc{MATLAB} environment, or through a pre-compiled binary, the latter of which does not require an installation of \textsc{MATLAB} to run.
A flow chart of the GUI is displayed in Fig.~\ref{fig:GenL_GUI}.
The \textsc{GenL} GUI consists of three different tabs corresponding to three steps: data import, simulating a sample, and fitting the data.
Alternatively, the user can also select an option in the import tab to simulate only without data import or fitting.
Following the first path, the user is asked to import the measurement which is to be fitted and prompted to enter the relevant measurement parameters.
In the simulate tab, the user supplies all relevant information on the expected crystal structure and layering of the sample as well as on the instrument setup.
In the fit tab, the user can proceed to tick a selection of parameters to be fitted.
Moreover, a diffraction pattern can at this stage be simulated based on the chosen parameter values and plotted against the imported data.
It is possible for the user to display the fit, along with the fitted parameters with their respective lower and upper bounds, the evolution of the figure of merit, and the fitted strain profile.

% ---------------
% ---------------

%************Application examples************
\section{Application examples}\vspace{0.5mm}

In this section we showcase some examples of employing \textsc{GenL} to fit diffraction patterns measured on W and Fe thin films.
Details on the thin film growth can be found in the respective publications for the W \cite{Ravensburg2023} and the Fe thin films \cite{Ravensburg2022, bct_Fe_Anna}.
All diffractograms presented in this section were measured in a Bede D1 diffractometer equipped with a Cu $K_{\alpha_1}$ X-ray source operated at 35~mA and 50~kV.
A circular mask (diameter: 0.005~m) and an incidence and a detector slit (both 0.0005~m) were used.
The beam was monochromatized by reducing the CuK$\beta$ and CuK$\alpha_2$ radiation using a G\"obel mirror and a 2-bounce-crystal on the incidence side.
The X-rays were detected with a Bede EDRc X-ray detector.
Details on growth and diffraction study of a Fe/V superlattice with a ratio of 4 monolayers of Fe to 28 monolayers of V in 11 bilayer repetitions, which is presented in comparison to a calculation with \textsc{GenL}, can be found elsewhere \cite{Droulias2017}.

\subsection{The influence of roughness}

\begin{figure}
\centering
\includegraphics[width=0.95\linewidth]{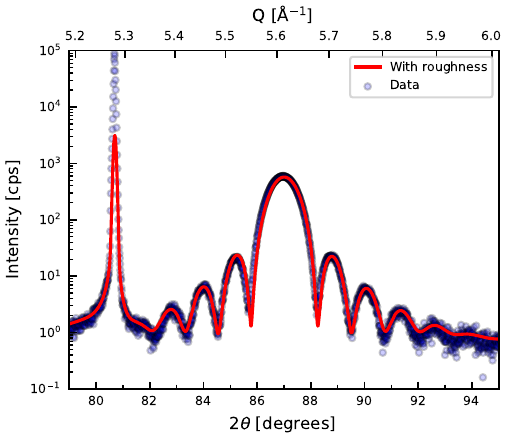}
\caption{X-ray diffraction pattern around the W~($220$) Bragg peak of a 100~{\AA} thick W layer grown on an Al$_{2}$O$_{3}$~($11\bar{2}0$) substrate. A fit of the data with \textsc{GenL} taking roughness into account is shown in red. The sharp peak at around 80~degrees originates from the single crystalline substrate.} 
\label{fig:rough}\hfill
\end{figure}

Thin W~($110$) layers can be epitaxially grown on Al$_{2}$O$_{3}$~($11\bar{2}0$) substrates with high crystal quality despite a large lattice mismatch of 7.2~\% and 19.4~\% along W~[$1\bar{1}1$] and W~[$\bar{1}12$] directions \cite{Ravensburg2023}.
The diffraction pattern of such a thin film, as shown in Fig.~\ref{fig:rough}, exhibits Laue oscillations around the W~($220$) Bragg peak.
The Laue oscillations have a high degree of symmetry indicating a small degree of strain in the layer.
Fitting the oscillations including roughness in \textsc{GenL} yields the fit displayed in red in Fig.~\ref{fig:rough}.
Based on the fit, the roughness is estimated to be 4~{\AA}.
This interface roughness is in agreement with roughness values at the W/capping interface obtained from Kiessig fringe fitting using \textsc{GenX} \cite{Bjorck2007, Glavic_Bjorck_2022} of the same sample \cite{Ravensburg2023}.
The analysis of the \textsc{GenL} fitting employed in \cite{Ravensburg2023} contributed to the understanding that W exhibits a semicoherent interface on Al$_{2}$O$_{3}$~($11\bar{2}0$) with an immediate strain relaxation taking place directly at the substrate/W interface and serves as an example of the successful application of \textsc{GenL} in studies on epitaxial growth.

% ---------------
% ---------------

\subsection{Accounting for overlapping substrate peak intensity}

\begin{figure}
\centering
\includegraphics[width=0.95\linewidth]{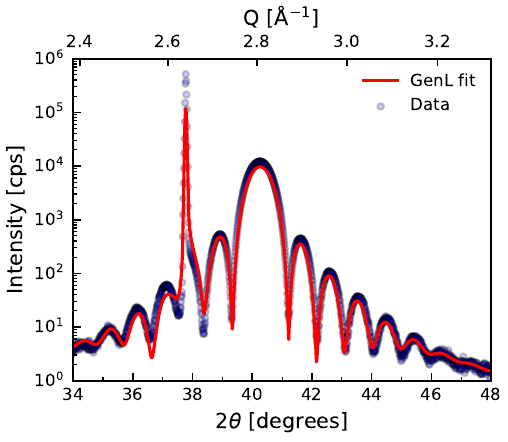}
\caption{X-ray diffraction pattern around the W~($110$) Bragg peak of a 100~{\AA} thick W layer grown on an Al$_{2}$O$_{3}$~($11\bar{2}0$) substrate. A fit of the data with \textsc{GenL} taking roughness into account as well as modeling intensity originating from the substrate, is shown in red.} 
\label{fig:substrate}\hfill
\end{figure}

A similar analysis can also be done with the scattering pattern around the W~($110$) Bragg peak instead, which is displayed in Fig.~\ref{fig:substrate}.
However, the sharp peak at around 38~degrees originating from the single crystalline Al$_{2}$O$_{3}$~($11\bar{2}0$) substrate overlaps in intensity with the Laue oscillations, making an analysis more challenging.
This angular overlap in the scattering patterns originating from film and substrate is common for epitaxial growth as it is attributed to the similar out-of-plane lattice parameters in the substrate and the film.
An example of a fit including roughness together with this modeled substrate peak intensity is displayed in red.
Both fits from Fig. \ref{fig:rough} and Fig. \ref{fig:substrate} for the same sample, yield similar fitted parameters.
The deviation of the $d_{\rm 110}$ obtained from both fits is 0.05~\%.

% ---------------
% ---------------

\subsection{The influence of strain}

\begin{figure}
\centering
\includegraphics[width=0.95\linewidth]{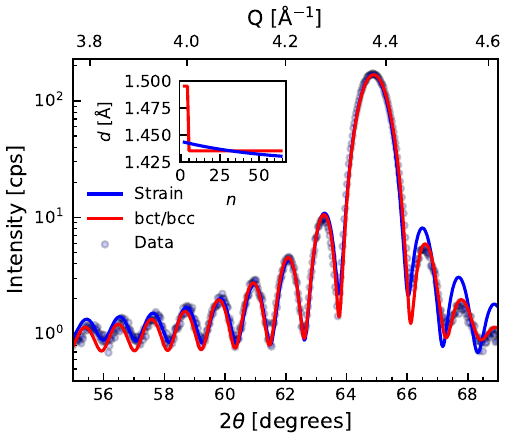}
\caption{X-ray diffraction pattern around the Fe~($002$) Bragg peak of a 100~{\AA} thick Fe layer grown on a MgAl$_{2}$O$_{4}$~($001$) substrate. Fits of the data with \textsc{GenL} taking tensile out-of-plane strain and a bct Fe/bcc Fe bilayer layering into account are shown in blue and red, respectively. The fitted strain profiles, i.e., the evolution of the interplanar spacing $d$ over the number of atomic layers $n$ are displayed in the inset in the respective color.} 
\label{fig:strain}\hfill
\end{figure}

Besides roughness and substrate peak intensity, \textsc{GenL} is also capable of taking a strain profile, i.e., a change in out-of-plane interplanar spacing over layer thickness into account, similar to \cite{Lichtensteiger2018}.
In Fig.~\ref{fig:strain} the diffraction pattern of a 100~{\AA} thick Fe layer grown on a MgAl$_{2}$O$_{4}$~($001$) substrate is displayed.
The \textsc{GenL} fit, employing an exponentially decaying strain profile, is shown in blue.
In the employed strain profile, the out-of-plane lattice spacing decreases over Fe layer thickness, in line with a tensile out-of-plane strain induced at the substrate/Fe interface by a lattice mismatch \cite{Ravensburg2022}.
The shape of this strain profile is in agreement with the observation that thinner Fe films, which are epitaxially grown on MgAl$_{2}$O$_{4}$~($001$), exhibit on average larger out-of-plane lattice spacings in the order of the predicted values in the profile.
However, the fit does not fully capture the asymmetric decay of the oscillations on both sides of the Bragg peak.
In an experimental study on thin Fe layers (6 to 100~{\AA}) grown on MgAl$_{2}$O$_{4}$~($001$) substrates \cite{bct_Fe_Anna}, Fe grows in a tetragonally distorted, body centered tetragonal (bct) crystal structure for Fe layer thicknesses below 3 to 8~ML before the equilibrium body centered cubic (bcc) crystal structure stabilizes on this substrate.
\textsc{GenL} can be used to investigate this growth even at the thin film limit.
Hence, the diffraction pattern was fitted employing the out-of-plane interplanar distance profile of a bct Fe/bcc Fe bilayer with both bilayer thicknesses and interplanar spacings as fitting parameters \cite{bct_Fe_Anna}.
The results are displayed in red in Fig.~\ref{fig:strain}.
It is evident that the fit based on the bilayer profile captures the detected scattering intensity better and reproduces the  asymmetry present in the intensity of the oscillations around the Fe~($002$) Bragg peak.
This example demonstrates the full potential of the application of \textsc{GenL} in research work.

% ---------------
% ---------------

\subsection{Limitations and possible extensions of the program}

Besides the factors already included in \textsc{GenL}, there are others that may also influence the measured diffracted intensity from an epitaxial thin film, e.g., sample alignment, the X-ray beam footprint, crystallographic defects, or mosaicity, all of which are not yet included in the program.
Some of these factors will be added in future versions of \textsc{GenL}.

\begin{figure}
\centering
\includegraphics[width=0.95\linewidth]{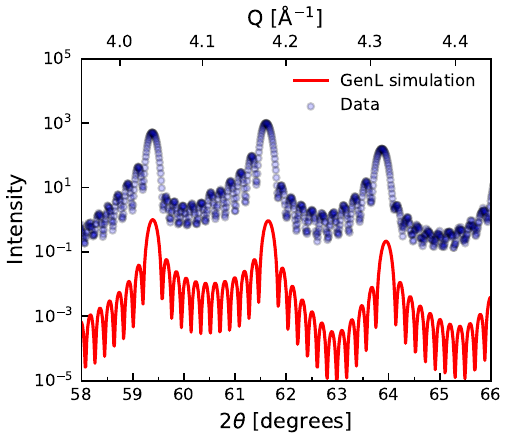}
\caption{X-ray scattering pattern from a Fe/V - 4/28 superlattice  with 11 bilayer repetitions \cite{Droulias2017} (blue) and a corresponding simulation of the diffraction intensity using \textsc{GenL} (red). The curves have been vertically shifted for clarity.} 
\label{fig:SL}\hfill
\end{figure}

The command line version of \textsc{GenL} can be extended to correctly calculate diffraction patterns for alloys due to the fact that \textsc{GenL} is modeling an effective one-dimensional stack of atoms whose order and positions can be adjusted. Even different/additional layers with multiple repetitions can be included.
Hence, \textsc{GenL} can be easily adapted for more complex multilayer stacks composed of different materials and/or different crystal structures, such as superlattice structures.
An example of a diffraction pattern from a Fe/V superlattice simulated in \textsc{GenL}, is displayed in Fig.~\ref{fig:SL} and compared against experimentally measured diffraction data \cite{Droulias2017}.

Fitting of superlattice diffraction data is not yet implemented in the current version of \textsc{GenL}, but will be included in future versions. The simulation of superlattice diffraction patterns is, however, straight forward in the command line version.
Moreover, the code versatility also enables the user to conduct a more in-depth analysis by simulating the presence of atomic step terraces.
In particular for epitaxial systems where out-of-plane atomic steps in the substrate are incommensurate with the out-of-plane atomic distance in the epitaxially grown film, interference of the scattering intensity from different stacks with vertical mismatch can contribute to an observed asymmetry in Laue oscillations \cite{Fullerton1992, Miller2022}, which may be relevant for fitting certain scattering patterns.

% ---------------
% ---------------

%************Conclusions************
\section{Conclusions}\vspace{0.5mm}

We developed and present a new versatile program, \textsc{GenL}, for the simulation of diffraction patterns from epitaxial thin films. The combination of this program along other existing X-ray reflectivity fitting programs, such as \textsc{GenX} \cite{Bjorck2007, Glavic_Bjorck_2022}, can form a powerful toolbox for rendering a fuller and more detailed structural picture for highly crystalline epitaxial thin films, exploiting X-ray scattering techniques. Finally, the ability to include and fit against experimental data, a series of interface (strain) and structural parameters (atomic steps, superlattices etc) will hopefully be a useful asset for the research community, encouraging a detailed and quantitative analysis of diffraction patterns comprising Laue oscillations.

%************Program distribution************
\section{Program distribution}\vspace{0.5mm}

\textsc{GenL} is freely available from the authors under the GNU General Public Licence (GPL).
A web page (URL: https://github.com/scatterer/GenL) exists for the distribution of the program.
The package contains the \textsc{Matlab} code, an executable of the GUI, example files for fitting and examples for sample simulations if the GUI is used. Comments, feedback, and support can be communicated using the following email addresses: {\it vassilios.kapaklis@physics.uu.se} and {\it gunnar.palsson@physics.uu.se}.

%-------------------------------------------------------------------------
% The back matter of the paper
%-------------------------------------------------------------------------
\vspace{6pt}
\noindent
{\bf Acknowledgements:} VK would like to acknowledge financial support from the Swedish Research Council (Project No. 2019-03581). GKP acknowledges funding from the Swedish Research Council (Project No. 2018-05200). The authors would like to express their gratitude to Prof. Dr. Gerhard Jakob for supplying the data presented in Fig.~\ref{fig:CADEM}.

\vspace{6pt}
\noindent
{\bf Data availability:} The data that support the findings of this study are available from the authors upon reasonable request.

%-------------------------------------------------------------------------
% References
%-------------------------------------------------------------------------

%\bibliographystyle{iucr}
%\bibliography{Fitting}

\providecommand{\noopsort}[1]{}\providecommand{\singleletter}[1]{#1}%

\end{document}